\documentclass[final,3p,times,twocolumn]{elsarticle}
\usepackage[hidelinks]{hyperref}
\usepackage{comment}
\usepackage{booktabs}
\usepackage{multirow}

\usepackage{url}
 
\makeatletter
\gdef\emailauthor#1#2{\stepcounter{ead}%
     \g@addto@macro\@elseads{\raggedright%
      \let\corref\@gobble\def\@@tmp{#1}%
      \eadsep{\normalfont\expandafter\strip@prefix\meaning\@@tmp}%
      (#2)\def\eadsep{\unskip,\space}}%
}
\gdef\urlauthor#1#2{\g@addto@macro\@elsuads{\let\corref\@gobble%
     \def\@@tmp{#1}\raggedright\eadsep
     {\normalfont\expandafter\strip@prefix\meaning\@@tmp}\space(#2)%
     \def\eadsep{\unskip,\space}}%
}
\makeatother

\usepackage{amssymb}
\usepackage{amsmath}
\journal{}

\begin{document}

\begin{frontmatter}

%% Title, authors and addresses

%% use the tnoteref command within \title for footnotes;
%% use the tnotetext command for theassociated footnote;
%% use the fnref command within \author or \affiliation for footnotes;
%% use the fntext command for theassociated footnote;
%% use the corref command within \author for corresponding author footnotes;
%% use the cortext command for theassociated footnote;
%% use the ead command for the email address,
%% and the form \ead[url] for the home page:
%% \title{Title\tnoteref{label1}}
%% \tnotetext[label1]{}
%% \author{Name\corref{cor1}\fnref{label2}}
%% \ead{email address}
%% \ead[url]{home page}
%% \fntext[label2]{}
%% \cortext[cor1]{}
%% \affiliation{organization={},
%%             addressline={},
%%             city={},
%%             postcode={},
%%             state={},
%%             country={}}
%% \fntext[label3]{}

\title{From Twelve Leads to Single: Cross-Lead Knowledge Distillation for Domain-Generalized Atrial Fibrillation Detection\tnoteref{t1}}

\tnotetext[t1]{This research was supported by Beijing Hospitals Authority Clinical Medicine Development of Special Funding Support [ZLRK202518] and Institute for Intelligent Healthcare, Tsinghua University [2023ZLB001].}

%% use optional labels to link authors explicitly to addresses:
%% \author[label1,label2]{}
%% \affiliation[label1]{organization={},
%%             addressline={},
%%             city={},
%%             postcode={},
%%             state={},
%%             country={}}
%%
%% \affiliation[label2]{organization={},
%%             addressline={},
%%             city={},
%%             postcode={},
%%             state={},
%%             country={}}

\author[1,2]{Xiaojian Lin\fnref{label2}}
\author[2]{Yang Li\fnref{label2}}
\author[3]{Yaomin Shen}
\author[4]{Wenxin Zhang}
\author[5]{Yu Liu}
\author[6]{Ping Zhang\corref{cor1}}
\ead{zhpdoc@126.com}
\author[7,2]{Bo Chen\corref{cor1}} 
\ead{cba04570@btch.edu.cn}
\cortext[cor1]{Equal corresponding authors}
\fntext[label2]{Equal contribution}

%% Author affiliation
\affiliation[1]{organization={Department of Biomedical Engineering, Tsinghua University, Beijing},%Department and Organization
        % addressline={}, 
        %    city={},
         %   postcode={}, 
        %    state={},
            country={China}}

\affiliation[2]{organization={Department of Electronic Engineering, Tsinghua University, Beijing},%Department and Organization
        % addressline={}, 
        %    city={},
         %   postcode={}, 
        %    state={},
            country={China}}

\affiliation[3]{organization={XR System Application R\&D Center, Nanchang Research Institute of Zhejiang University, Nanchang 330096},%Department and Organization
        % addressline={}, 
        %    city={},
         %   postcode={}, 
        %    state={},
            country={China}}

\affiliation[4]{organization={School of Computer Science of Technology, University of Chinese Academy of Sciences,, Beijing 101408},%Department and Organization
        % addressline={}, 
        %    city={},
         %   postcode={}, 
        %    state={},
            country={China}}

\affiliation[5]{organization={Zhuhai People's Hospital, Zhuhai},%Department and Organization
        % addressline={}, 
        %    city={},
         %   postcode={}, 
        %    state={},
            country={China}}

\affiliation[6]{organization={Beijing Tsinghua Changgung Hospital, Beijing},%Department and Organization
        % addressline={}, 
        %    city={},
         %   postcode={}, 
        %    state={},
            country={China}}
            
\affiliation[7]{organization={School of Clinical Medicine, Tsinghua University, Beijing},%Department and Organization
        % addressline={}, 
        %    city={},
         %   postcode={}, 
        %    state={},
            country={China}}

\begin{abstract} 
Atrial fibrillation (AF) detection from long-term ambulatory electrocardiography (ECG) is critical because AF is a leading cause of stroke yet is frequently missed by conventional diagnostics, particularly in wearable settings that demand strong cross-domain generalization. Deep learning has improved AF identification by exploiting rhythm patterns, but methods such as BMIRC and SJTU-ECGNet still struggle to generalize across heterogeneous single-lead wearable datasets. To address this issue, we propose X-LeadGen, a two-stage cross-lead distillation framework for single-lead ECG AF detection that explicitly targets domain generalization. First, a 12-lead teacher network distills rich multi-lead information into a single-lead student through cross-lead knowledge transfer and a simulated self-distillation branch within the teacher, improving cross-lead representation alignment and AF discrimination. Second, we introduce a domain-generalization strategy that applies domain-adversarial training in both teacher and student stages, complemented by GANs-based augmentation in the student phase to learn domain-invariant features and enhance robustness to distribution shifts. Across seven public ECG datasets under real and simulated Lead-I settings, X-LeadGen consistently outperforms existing approaches in accuracy and F1-score, demonstrating strong generalization for real-world wearable AF monitoring. Our implementation is publicly available at \href{https://github.com/Aveiro-Lin/X-LeadGen}{https://github.com/Aveiro-Lin/X-LeadGen}
, enabling further research and validation in the ECG community.

\end{abstract}

%% Keywords
\begin{keyword}
Single-Lead ECG\sep Domain Generalization\sep AF Detection.
\end{keyword}
\end{frontmatter}

%% Add \usepackage{lineno} before \begin{document} and uncomment 
%% following line to enable line numbers
%% \linenumbers

%% main text
%%

\section{Introduction}

\label{sec:introduction}

In recent years, the rapid proliferation of wearable devices such as the Apple iWatch\cite{Johnson-2025} has made single-lead electrocardiography (ECG)\cite{Zepeda-2023} an indispensable modality for home-based and community healthcare. Long-term ambulatory single-lead recordings are now routinely acquired outside the hospital, generating massive volumes of ECG data that can, in principle, support continuous monitoring, chronic disease management, and timely arrhythmia detection. However, in real-world deployments, patients are widely distributed, devices differ substantially in hardware and signal processing pipelines, and annotated data remain scarce. These factors induce pronounced variability in data distribution, signal quality, and label availability, placing stringent requirements on the generalization capability of machine learning models\cite{Wang-2024}.

State-of-the-art deep learning approaches have demonstrated that, under controlled conditions, multi-lead ECG classification models can achieve very high diagnostic accuracy—often exceeding 90\%\cite{Bai-2024}. Yet such performance does not reliably transfer across heterogeneous domains\cite{Ding-2025}. The problem is further exacerbated in single-lead settings: the reduced spatial information makes it intrinsically harder to capture comprehensive cardiac activation patterns\cite{BenMoshe-2024,Attia}, and models trained on a specific dataset tend to overfit its idiosyncrasies. As a result, single-lead models frequently exhibit strong in-distribution performance but degrade sharply when confronted with unseen devices, institutions, or patient populations\cite{BenMoshe-2024,Wang-2024}, undermining their reliability for large-scale wearable deployment.

From a modeling perspective, current strategies are ill-suited to bridge this gap between powerful multi-lead models and deployment-ready single-lead solutions\cite{An-2024}. High-capacity multi-lead networks\cite{Bai-2024} exploit rich spatial context but are typically too resource-intensive for wearable or low-power platforms, and they operate under lead configurations that are fundamentally different from those of consumer devices. Single-lead models, while computationally lightweight and deployment-friendly, are notoriously sensitive to cross-dataset distribution shifts\cite{Robey,Huang-2024}, precisely because they must infer complex rhythm and morphology patterns from severely limited input dimensionality\cite{Wang-2024}. Consequently, neither class of models, in its current form, adequately addresses the robustness and generalization requirements of real-world, single-lead wearable ECG analysis.

At the same time, existing efforts to improve generalization remain fragmented. Techniques such as domain-invariant representation learning, adversarial training, and self-supervised objectives have been explored in ECG and related biosignal domains\cite{Benchaira-2024,Xie-2024,Islam2024,Na2024}, but they are often developed in isolation and rarely integrated into a coherent, deployment-oriented framework\cite{Jin2024}. In particular, current methods seldom exploit the rich diagnostic information embedded in multi-lead clinical ECG to systematically guide the learning of single-lead models\cite{Na2024}, nor do they jointly tackle lead-configuration mismatch and multi-domain distribution shift in a unified manner\cite{Wei2025BMIRC}. This fragmentation leaves a critical methodological gap for robust, scalable single-lead ECG analysis in wearable scenarios.

To address the aforementioned limitations and improve the robustness and generalizability of single-lead ECG classification models, we propose X-LeadGen, a two-stage cross-lead knowledge transfer framework specifically tailored for atrial fibrillation (AF) detection in wearable ECG data. Our main contributions are as follows:

\begin{itemize}

\item We propose X-LeadGen, a two-stage cross-lead distillation framework that transfers rich 12-lead knowledge to single-lead AF detection via a novel simulated self-distillation mechanism in the teacher.
\item We enhance domain generalization by integrating domain-adversarial training in both teacher and student stages, further combined with GANs-based augmentation in the student phase.
\item Extensive experiments on seven public ECG datasets under real and simulated single-lead settings show that X-LeadGen significantly outperforms state-of-the-art methods (e.g., BMIRC\cite{Wei2025BMIRC} and SJTU-ECGNet\cite{Jin2024}) in accuracy, F1-score, and cross-dataset generalization.

\end{itemize}

The rest of our paper is structured as follows: 
In Sec.II, we provide a review of related works to contextualize our research. 
Next, in Sec.III, we explain each key component in our pipeline. 
Subsequently, in Sec. IV, we evaluate the performance of our method through various setup of ablation study. 
Finally, the limitation and  conclusion are drawn in Sec.V and VI.

\begin{figure*}[h]
\centering
\includegraphics[width=\linewidth]{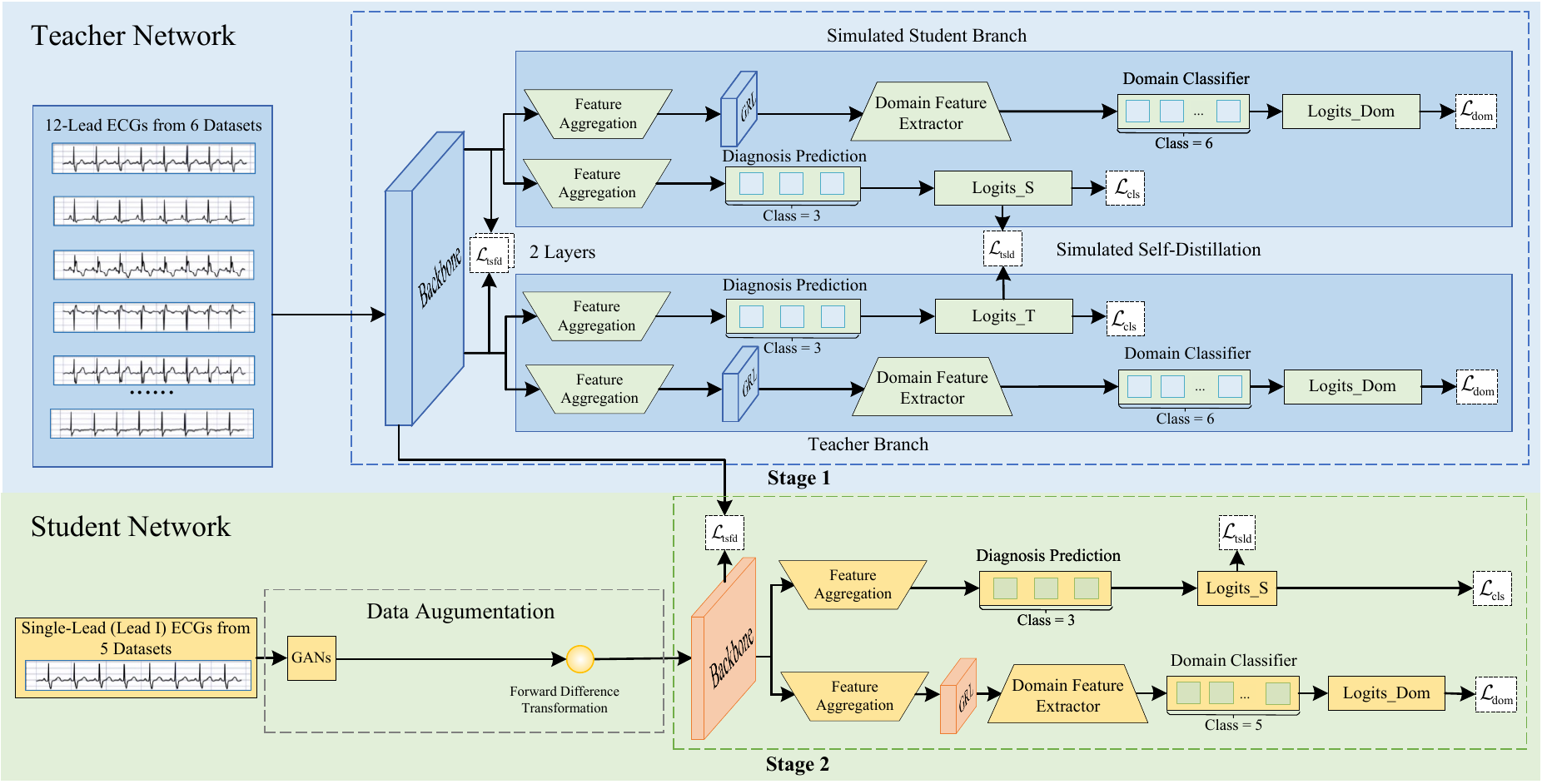}
\caption{The two-stage X-LeadGen pipeline for single-lead AF detection. (a) Stage 1 serves as the teacher network. Stage 1 trains a 12-lead teacher with domain-adversarial learning and transfers knowledge to a simulated single-lead student via simulated self-distillation. (b) Stage 2 is the student network. Stage 2 trains a single-lead student on protocol-defined source domains, with optional GANs-based augmentation and domain-adversarial learning for cross-dataset generalization. The displayed five-source Stage-2 setting corresponds to a leave-one-dataset-out student fold.
}
\label{fig:overview of architecture}
\end{figure*}

\section{Related Work}
\subsection{Single-Lead ECG Classification}
Single-lead ECG classification has gained significant attention due to its widespread application in wearable devices, offering a low-cost and accessible solution for continuous health monitoring. The primary focus in this area has been the development of models that can accurately classify arrhythmias and other heart conditions from a single-lead ECG signal. Several deep learning methods have been proposed for ECG classification, including 1D-CNN-based models\cite{TKASP-2020}, ResNet architectures\cite{QIN2023107503, Aristotelis Ballas_2022}, and Transformer models\cite{ElGhaish2024}, which have shown strong performance on various datasets.

Despite these advances, one of the main challenges in single-lead ECG classification is the performance variability across different devices and environments. Unlike multi-lead ECGs, single-lead signals capture limited information, making them highly sensitive to discrepancies in device types, electrode placement, and signal noise. To address this issue, domain adaptation and generalization methods have been explored to improve model robustness in the presence of such variations. However, these methods often fail to generalize well across different ECG devices and populations, which leads to significant performance degradation on unseen data. This limitation arises because they do not explicitly address the lead-configuration mismatch or the substantial distribution shifts inherent to wearable ECG acquisition.

To overcome these challenges, some works have emphasized improving the generalization capability of single-lead ECG models. Techniques like data augmentation, domain-adversarial training and meta-learning have been proposed to make the models more adaptable to different environments\cite{TKASP-2020, Aristotelis Ballas_2022, Aristotelis Ballas_2024}. Despite these efforts, a comprehensive solution that consistently improves the generalization of models across a wide range of devices and scenarios remains an open problem.
To address this gap, X-LeadGen introduces a unified framework that strengthens morphological representation, domain robustness, and signal-level adaptability. Through cross-lead knowledge transfer, two-stage domain-adversarial training, and GAN-based augmentation, it effectively mitigates distribution shifts and achieves more reliable generalization across heterogeneous devices and real-world wearable settings.

\subsection{Domain Generalization in Medical Signals}

Domain generalization aims to train models on one or more source domains such that they generalize well to unseen target domains with different data distributions. In medical signal analysis, this is particularly important due to high inter-patient variability, sensor heterogeneity, and domain shifts across devices and environments.

Classical domain generalization techniques have been introduced to alleviate domain shift, including:\textit{1. Maximum Mean Discrepancy (MMD)}: minimizes the distribution distance between domains in a reproducing kernel Hilbert space~\cite{Gretton2012}. \textit{2. Gradient Reversal Layer (GRL)}: encourages domain-invariant feature extraction via domain-adversarial training~\cite{Ganin}. \textit{3. Invariant Risk Minimization (IRM)}: promotes predictors to learn features that remain invariant across multiple environments~\cite{Arjovsky2019}. While these methods work well in general ML and some biomedical tasks, their performance on structured signals like ECG is still limited: they rarely transfer \textit{structured dependencies} in physiological data (e.g., inter-lead relations or temporal rhythm patterns) and typically act only on latent feature distributions without explicit \textit{physiological structure awareness}. Even recent ECG domain generalization studies using domain-adversarial or multi-source learning~\cite{Aristotelis Ballas_2024,Jiyao Wang_2024} leave such structural modeling largely underexplored.

In contrast, X-LeadGen incorporates physiological structure through cross-lead knowledge transfer, enabling the student to inherit inter-lead morphological dependencies that classical domain generalization methods miss. Combined with feature/logit distillation and domain-adversarial objectives, it aligns both physiological representations and domain distributions, achieving stronger generalization on unseen single-lead ECG domains.

\subsection{Knowledge Distillation and Cross-Lead Transfer}

Knowledge distillation (KD) is a widely used technique in model compression and representation transfer, where a smaller student model is trained to mimic the behavior of a larger, pre-trained teacher model~\cite{Waheed2025}.
Classical KD approaches include matching the teacher’s output logits or intermediate feature representations during student training~\cite{Cheng2025EEG2IM,Wang2025MMKD,Su2025Osteo}.
These methods have demonstrated effectiveness across various domains, including medical signal analysis~\cite{Wei-2025}.

In the context of ECG, recent works have explored lead-specific knowledge transfer, aiming to bridge the representational gap between multi-lead and single-lead ECG models. For instance, the MVKT-ECG method~\cite{QIN2023107503} employs contrastive learning to enhance the alignment of disease-relevant features across leads. It transfers global rhythm and morphological patterns from multi-lead representations to single-lead models. However, most of these approaches focus on aligning feature spaces or logits, lacking an explicit structural formulation for lead-wise knowledge flow or self-enhancement.

In contrast, our method is the first to introduce a structurally organized cross-lead knowledge transfer framework that not only distills knowledge from a multi-lead teacher to a single-lead student, but also enhances the teacher model itself through a novel simulated self-distillation mechanism. This formulation enables the teacher to internally refine its feature representations by learning from a simulated student branch, a design not considered in prior works.

\section{Method}
\subsection{Overview}

\noindent\textbf{Task Description.} The detection of AF via AI models represents a critical task in the field of medical engineering, given its potential to enhance early diagnosis and improve patient outcomes. In this work, we introduce X-LeadGen, a novel AF detection method tailored for single-lead electrocardiogram (ECG) data, optimizing its applicability in wearable device environments for continuous monitoring.

\noindent\textbf{Overall Framework.} As shown in Fig.\ref{fig:overview of architecture}, X-LeadGen consists of two primary stages. The \textbf{Stage 1} leverages a multi-lead ECG architecture, enhanced with domain-adversarial training in the teacher branch to extract domain-invariant features. This full-source pretraining stage processes 12-lead ECG data from 6 datasets to transfer knowledge to a simulated student, utilizing simulated self-distillation to improve feature representation. The \textbf{Stage 2} trains a single-lead student network on Lead-I signals from the source datasets defined by each evaluation protocol, while a frozen 12-lead teacher provides logits and feature embeddings for label- and feature-level distillation. The Stage-2 domain classifier therefore changes with the source set: six domains in the real wearable experiment and five domains in each leave-one-dataset-out student fold. When enabled, GANs-based data augmentation and forward difference transformation are applied to the input single-lead ECG data to improve generalization across diverse datasets. In our implementation, student optimization combines diagnosis loss, domain-adversarial loss, and teacher-guided distillation losses.

\label{subsection:overview}

\noindent\textbf{Backbone Design.} Both stages use the same backbone, differing in input dimensionality (12-lead vs. single-lead ECG). The architecture follows the 33-layer residual CNN of Hannun et al.~\cite{Hannun}, but replaces standard residual blocks with 16 Fibrillation Residual (FibRes) Blocks and a final Feature Aggregation module. Each FibRes Block applies max pooling \emph{before} convolution for early temporal abstraction, allowing convolutions to operate on compressed sequences and better capture long-range AF patterns. Skip connections are made after the second convolution, and channel mismatches are handled by zero-padding the identity branch every four blocks, stabilizing training and improving AF-specific temporal representations. A detailed schematic of the backbone and the data flow within each FibRes Block is provided in Supplementary Fig.~\ref{fig:supp_fibres_backbone}.

\begin{comment}

(Fig.~\ref{fig:backbone compare})

\begin{figure}[t]
\begin{center}
   \includegraphics[width=0.5\columnwidth]{backbone_compare.pdf}
\end{center}
\vspace{-10pt}
\caption{The architecture of the FibRes Backbone for AF detection in single-lead ECG. It consists of 16 stacked FibRes blocks, each featuring convolutional layers, batch normalization, max pooling, and a residual connection for enhanced feature learning and temporal abstraction.}
%\vspace{-18pt}
\label{fig:backbone compare}
\end{figure}

\end{comment}

\noindent\textbf{Feature Aggregation.} To improve temporal feature summarization, we modify the output head from Hannun et al.~\cite{Hannun}. Instead of directly mapping convolutional features to class logits, we make three key changes. First, an \textit{AdaptiveAvgPool1d} layer generates a fixed-length temporal embedding for global rhythm abstraction. Second, we replace \textit{ReLU} with \textit{LeakyReLU} to retain negative responses and enhance sensitivity to irregular AF components. Third, the pooled features are flattened into a vector ($x \in \mathbb{R}^{B \times 256}$), ensuring a well-conditioned input for downstream tasks. This approach improves robustness to signal variability and better supports rhythm-level predictions.

\noindent\textbf{Diagnosis Branch.} In both \textbf{Stage 1} and \textbf{Stage 2}, branches with a diagnosis prediction head are referred to as the Diagnosis Branch. This branch is shared by both teacher and student networks. Backbone features are summarized by the \textit{Feature Aggregation} module and passed into a classification head, which outputs probabilities over three rhythm categories: \textbf{AF}, \textbf{normal}, and \textbf{other}. The logits, \(\text{Logits}_T\) for the teacher and \(\text{Logits}_S\) for the student, are used for a classification loss \(\mathcal{L}_{\text{cls}}\) and a label-distillation loss \(\mathcal{L}_{\text{tsld}}\), encouraging the student to match the teacher’s predictive distribution. The teacher network also enables \emph{simulated self-distillation}, refining decision boundaries using its own logits as soft targets. These hierarchical distillation pathways are crucial for robust domain adaptation across diverse ECG devices and datasets.

\noindent\textbf{Data Augmentation Process.}
In \textbf{Stage 2}, when augmentation is enabled, we use a GAN-based augmentation pipeline to enrich the single-lead ECG training distribution.
Specifically, we employ a lightweight UNIT-style 1D generator--discriminator architecture, where the generator adopts an encoder--decoder design with AdaIN-based style injection ($\text{style\_dim}=8$) and the discriminator is implemented as a multi-layer 1D CNN operating on raw waveforms.
The GAN is trained \emph{offline} with Adam ($\text{lr}=1\times10^{-4}$, $\beta=(0.5,0.999)$) for $50$ epochs (batch size $1$); we log generator/discriminator losses and periodically visualize real vs.\ synthesized ECG segments as a sanity check to monitor training stability and detect obvious artifacts.
After training, the generator is frozen and only used to synthesize additional segments, which are mixed with real recordings to train the student network, increasing coverage of inter-patient variability and noise conditions (the final AF detector does not rely on the GAN at inference time).
The augmented signals are then transformed with a forward difference transformation, emphasizing temporal changes and rhythm dynamics critical for arrhythmia detection.
When GAN augmentation is disabled, we set $K_\mathrm{aug}=0$, reducing the student objective to the original-view case.
This augmentation--transformation scheme enhances data efficiency and generalization, enabling robust AF detection in low-resource wearable settings.

\label{subsection:BM}

\subsection{Teacher Network}

In the full-source teacher pretraining stage, 12-lead ECG recordings from six clinical datasets are first processed by the backbone to extract latent representations. These representations are then forwarded to both the teacher branch and the simulated-student branch. Prior to entering the two branches, a two-layer feature distillation module is applied to align and refine the extracted features.

\begin{comment}
The \textbf{cross-lead feature distillation} framework, which transfers rich 12-lead rhythm representations from the teacher to the single-lead student by minimizing their feature discrepancy, enabling the student to inherit multi-lead rhythm knowledge from compact single-lead inputs.
\end{comment}

\noindent\textbf{Teacher--Simulated-Student Feature Distillation.}
During the \textbf{teacher training stage}, the multi-lead backbone and the internal Lead-I simulated student branch perform \emph{intra-teacher feature distillation}. 
Unlike a single-layer distillation, our implementation introduces \emph{two} feature-level constraints, corresponding to two different depths of the feature hierarchy. 
Specifically, the teacher provides intermediate feature maps $\{F^{(12)}_1, F^{(12)}_2\}$, and the simulated student outputs $\{T^{(I)}_1, T^{(I)}_2\}$ at matched layers. 
The attention-based feature distillation loss is therefore defined as the sum of two attention transfer (AT) terms:
\begin{equation}
\label{eq:tsfd-teacher}
\mathcal{L}_{\text{tsfd}}^{12\leftrightarrow I}
= \lambda_{f_1}\,\mathcal{L}_{\text{AT}}\!\big(F^{(12)}_1,T^{(I)}_1\big)
+ \lambda_{f_2}\,\mathcal{L}_{\text{AT}}\!\big(F^{(12)}_2,T^{(I)}_2\big),
\end{equation}
where each $\mathcal{L}_{\text{AT}}$ measures the attention alignment between corresponding \emph{1D temporal} feature maps by first aggregating channel energy to form a length-$T$ attention vector and then computing the $\ell_2$ distance after $\ell_2$ normalization along the temporal axis (i.e., $A(U)=\mathrm{Norm}(\sum_c |U_{c,:}|^2)$, $\mathcal{L}_{\text{AT}}(U,V)=\|A(U)-A(V)\|_2^2$). 
This definition is \emph{parameter-free} and does not require matching channel dimensions; if the temporal lengths of a matched pair differ, we align them to the same resolution via 1D interpolation before computing $\mathcal{L}_{\text{AT}}$. 
Both weights $\lambda_{f_1}$ and $\lambda_{f_2}$ are set to $0.3$. 
These two layers are deliberately chosen: the shallow feature map $F^{(12)}_1$ preserves local waveform morphology, while the deeper map $F^{(12)}_2$ captures long-range rhythm dependencies. 
By enforcing consistency at both levels, the simulated student learns to approximate the teacher’s multi-scale representation of ECG dynamics, ensuring that knowledge about both morphology and temporal rhythm is preserved before aggregation.

After feature distillation, the distilled representations are passed into both the teacher branch and the simulated-student branch. These two branches share a similar structure, each consisting of a diagnosis branch and a domain-adversarial branch. The diagnosis branch has already been described in detail in the previous section, and we thus omit architectural repetition here. Instead, we focus on the logits produced by the two branches—denoted as Logits-T (teacher) and Logits-S (simulated student). Following diagnosis prediction, we perform label distillation between these two outputs. The label distillation framework, which governs the transfer of diagnostic knowledge at the logit level. Unlike formulations based on temperature-scaled KL divergence, our \emph{implemented} losses operate \emph{directly on logits} via cosine similarity, which we found to be numerically stable and faithful to the training code. We refer to this label distillation process as a form of simulated self-distillation.

\begin{comment}

In summary, the Teacher–Simulated-Student feature distillation employs \textbf{two} AT-based constraints ($F_1$ and $F_2$) to ensure multi-scale consistency within the teacher, while the teacher–student distillation in the second stage applies a \textbf{single} contrastive loss to bridge the cross-lead representation gap. 
Both mechanisms are independent of the domain-adversarial branch and operate before the feature aggregation modules, enabling X-LeadGen to achieve coherent, lead-invariant ECG representations that generalize effectively across domains.
\end{comment}

\noindent\textbf{Simulated Self-Distillation.} 
Let $\mathbf{z}_T^{(12)} \in \mathbb{R}^C$ and $\mathbf{z}_S^{(I)} \in \mathbb{R}^C$ denote the class logits (with $C{=}3$) from the 12-lead Teacher and the Lead-I Simulated-Student, respectively. 
The label-level distillation loss minimizes their discrepancy using a cosine objective:
\begin{equation}
\label{eq:tsld-teacher}
\mathcal{L}_{\text{tsld}}^{12\leftrightarrow I}
= \frac{1}{B}\sum_{b=1}^{B}
\Big(1 - \cos\!\big(\mathbf{z}_{T,b}^{(12)},\,\mathbf{z}_{S,b}^{(I)}\big)\Big),
\end{equation}
where $B$ is the batch size and 
$\cos(\mathbf{a},\mathbf{b})=\frac{\mathbf{a}^{\top}\mathbf{b}}{\|\mathbf{a}\|_2\,\|\mathbf{b}\|_2}$.
This bidirectional alignment reconciles global (multi-lead) and local (single-lead) decision representations inside the teacher. We adopt a cosine loss on logits for label-level distillation instead of the standard temperature-scaled KL divergence.
In our cross-domain domain generalization setting, different datasets/devices can induce non-negligible variations in logit scale and confidence calibration, which may make temperature-sensitive KL optimization less stable.
Cosine loss emphasizes the directional agreement of class evidence and is invariant to global logit scaling, thereby providing improved numerical stability under domain shifts.

\noindent\textbf{Domain-Adversarial Branch.} 
After introducing the simulated self-distillation mechanism in the diagnosis branch, we next describe the Domain-Adversarial Branch. This branch enables the network to learn \emph{domain-invariant representations} that generalize across heterogeneous ECG datasets and acquisition conditions. As illustrated in Fig.~\ref{fig:overview of architecture}, the aggregated features first pass through a lightweight \emph{Domain Feature Extractor}, which transforms them into a domain-sensitive embedding. This embedding is then input to a multi-layer \emph{Domain Classifier} that outputs the domain logits, one for each dataset source. Let $C_{\mathrm{src}}$ denote the number of source domains in the current training protocol: $C_{\mathrm{src}}=6$ for full-source teacher pretraining and real wearable training, and $C_{\mathrm{src}}=5$ in each leave-one-dataset-out student fold. Consistent with the classical \textit{Gradient Reversal Layer (GRL)} paradigm~\cite{Ganin}, the domain classifier is optimized using standard cross-entropy to correctly identify the domain, whereas the feature extractor receives a sign-reversed gradient. This adversarial interaction encourages the upstream backbone to produce domain-invariant representations, thereby reducing the model’s sensitivity to inter-dataset variability.

In the teacher network, both the 12-lead Teacher and the Lead-I Simulated-Student branch are equipped with independent domain classifiers, $D^{(12)}$ and $D^{(I)}$, respectively. 
Given feature representations $F^{(12)}(x)$ and $F^{(I)}(x)$, and corresponding domain labels $g$, the domain-adversarial loss is formulated as:
\begin{equation}
\label{eq:dom-teacher-ce}
\begin{aligned}
\mathcal{L}^{T}_{\mathrm{Dom}}
&=\frac{1}{B}\sum_{b=1}^{B}\!
\Big[
\mathrm{CE}\!\big(D^{(12)}(F^{(12)}(x_b)), g_b\big) \\
&\quad
+\,\mathrm{CE}\!\big(D^{(I)}(F^{(I)}(x_b)), g_b\big)
\Big],
\end{aligned}
\end{equation}

where $\mathrm{CE}(\cdot)$ denotes the cross-entropy loss. 
The GRL inverts the gradient only for the feature extractors, ensuring that both branches learn features that are discriminative for AF diagnosis yet indistinguishable across domains. 
This design allows the teacher to suppress dataset-specific bias at both the global (multi-lead) and local (single-lead) representation levels.

\subsection{Student Network}
%\vspace{-5pt}
After establishing the design and training paradigm of the Stage-1 teacher network, we now introduce the Stage-2 student network. The student is trained on the source datasets specified by each evaluation setting: for the real wearable test, all six 12-lead datasets provide source-domain Lead-I signals, whereas for leave-one-dataset-out student evaluation the held-out dataset is excluded and the remaining five datasets are used for student training and validation. The domain-classifier output dimension is adjusted accordingly, preventing target-domain leakage in Stage~2 while preserving a consistent single-lead training formulation across experiments.

\noindent\textbf{Teacher--Student Feature Distillation.}
Before extracting the training samples from the source datasets, we apply data augmentation to the single-lead (Lead-I) signals in the student stage when enabled. This procedure has already been described in detail in the Overview section and is therefore not reiterated here. The resulting signals are then fed into the backbone for feature extraction, where our core innovation—feature distillation—is introduced. 

While the distillation mechanism follows the general structure used in the teacher stage, several key differences remain in its formulation and application.
The pre-trained teacher is frozen and serves as the knowledge source for the external single-lead student. 
Given a batch of $1+K_\mathrm{aug}$ views (one original sample and $K_\mathrm{aug}$ GANs-augmented variants), we compute a single contrastive feature distillation loss $\mathcal{L}_{\text{tsfd}}^{T\rightarrow S}$ for each view and average across them:
\begin{equation}
\label{eq:tsfd-student}
\begin{aligned}
\tilde{\mathcal{L}}_{\mathrm{tsfd}}^{T\to S}
&=\frac{1}{(1+K_\mathrm{aug})N}
\sum_{i=0}^{K_\mathrm{aug}}\sum_{n=1}^{N}\\
&\quad
\max\!\big(0, s_i^{-}(n)-s_i^{+}(n)+\Delta\big),
\end{aligned}
\end{equation}

where $s_i^\pm(n)=\mathrm{sim}\!\big(\mathbf{F}_S^{(i)}(n),\mathbf{F}_T^\pm(n)\big)$, $\mathrm{sim}(\cdot,\cdot)$ denotes cosine similarity, and $N$ is the number of feature instances used in the contrastive term. $\mathbf{F}_S^{(i)}$ denotes the student features from the $i$-th view, and $\mathbf{F}_T^{+},\mathbf{F}_T^{-}$ are positive and negative teacher features sampled from the maintained feature queue. We set the margin $\Delta$ to $0.1$ in all experiments. Unlike the two-layer intra-teacher alignment, the inter-model feature distillation uses a single-level embedding after feature aggregation, since the student receives only one lead and thus lacks the hierarchical multi-scale structure of the teacher. 
This margin loss enforces the student feature to be more similar to positive teacher features than to negative teacher features by at least $\Delta$, aligning the global rhythm representation between teacher and student.

\noindent\textbf{Label Distillation.} 
After feature distillation, the resulting representations are routed into two branches whose structures closely mirror the simulated-student and teacher branches in Stage 1. These two branches consist of a diagnosis head and a domain-adversarial head, and their architectural details are therefore omitted here. Instead, we highlight the key differences relative to Stage 1. In the diagnosis branch, the student’s predicted logits (Logits-S) are further subjected to label distillation—analogous to the simulated self-distillation performed in Stage 1. 

During the \textbf{student training stage}, the trained teacher is frozen and provides soft supervision to the single-lead student. 
Given teacher logits $\mathbf{z}_T$ and student logits $\mathbf{z}_S$ on the same input, the inter-model label distillation is
\begin{equation}
\label{eq:tsld-student-single}
\mathcal{L}_{\text{tsld}}^{T\rightarrow S}
= \frac{1}{B}\sum_{b=1}^{B}
\Big(1 - \cos\!\big(\mathbf{z}_{T,b},\,\mathbf{z}_{S,b}\big)\Big).
\end{equation}
When $K_\mathrm{aug}$ GANs-augmented variants are generated per batch in addition to the original sample ($i{=}0$ for original and $i{=}1,\ldots,K_\mathrm{aug}$ for GANs variants), we aggregate the distillation over all views:
\begin{equation}
\label{eq:tsld-student-gan}
\widetilde{\mathcal{L}}_{\text{tsld}}^{T\rightarrow S}
= \frac{1}{1+K_\mathrm{aug}}\sum_{i=0}^{K_\mathrm{aug}}
\frac{1}{B}\sum_{b=1}^{B}
\Big(1 - \cos\!\big(\mathbf{z}^{(i)}_{T,b},\,\mathbf{z}^{(i)}_{S,b}\big)\Big).
\end{equation}

\noindent\textbf{Domain-Adversarial Branch.} 
In the domain-adversarial branch, most components remain analogous to those in Stage 1; however, the incorporation of GAN-based augmentation introduces several differences in the formulation of the loss. The student model contains a single domain classifier $D^{(S)}$ attached after its feature extractor $F_S$. 
In each training batch, we consider the original single-lead input ($i=0$) and $K_\mathrm{aug}$ GANs-augmented variants ($i=1,\ldots,K_\mathrm{aug}$). 
For each view, the GRL operates identically to the teacher case, and the corresponding domain-adversarial loss is averaged across all augmented views:
\begin{equation}
\label{eq:dom-student-ce}
\begin{aligned}
\mathcal{L}^{S}_{\mathrm{Dom}}
&=\frac{1}{(1+K_\mathrm{aug})B}
\sum_{i=0}^{K_\mathrm{aug}}\sum_{b=1}^{B} \\
&\quad
\mathrm{CE}\!\big(D^{(S)}(F_S^{(i)}(x_b)), g_b\big).
\end{aligned}
\end{equation}
This setup encourages the student to maintain domain-invariant behavior even under augmented and distribution-shifted single-lead inputs.

\subsection{Loss Function}
%\vspace{-5pt}

The optimization of X-LeadGen is formulated through two hierarchical objectives that exactly follow the implementation. 
Each stage combines diagnostic supervision, domain-adversarial regularization, and knowledge distillation, with explicit weights tuned to maintain an effective balance between stability and transfer efficiency.

\noindent\textbf{Stage 1.}
During teacher training, the 12-lead Teacher and the internal Lead-I Simulated-Student branch are optimized jointly. 
The total objective integrates diagnostic classification, dual GRL-based domain alignment, and multi-level distillation losses as follows:
\begin{equation}
\label{eq:teacher-total-loss-final}
\begin{aligned}
\mathcal{L}_{\text{Teacher}}
&=
\underbrace{\mathcal{L}_{\text{cls}}^{(12)} + \mathcal{L}_{\text{cls}}^{(I)}}_{\text{diagnostic classification}} 
+ \underbrace{0.05\,\mathcal{L}_{\text{Dom}}^{T}}_{\text{domain-adversarial branches}} \\
&\quad
+ \underbrace{0.3\,\mathcal{L}_{\text{tsld}}^{12\leftrightarrow I}}_{\text{label distillation}}
+ \underbrace{\mathcal{L}_{\text{tsfd}}^{12\leftrightarrow I}}_{\text{feature distillation}}.
\end{aligned}
\end{equation}

Here, the diagnostic classification loss $\mathcal{L}_{\text{cls}}^{(12)}$ and $\mathcal{L}_{\text{cls}}^{(I)}$ are standard cross-entropy objectives with implicit unit weights (1.0), ensuring that the backbone and its Simulated-Student branch both retain strong discriminative capacity for AF, normal, and other rhythm categories. 
The GRL weight $0.05$ stabilizes domain adaptation by introducing sufficient adversarial pressure without overwhelming the classification objective, while the distillation weights $0.3$ act as moderate regularizers—strong enough to align representations but small enough to preserve discriminability. 
$\mathcal{L}_{\text{tsfd}}^{12\leftrightarrow I}$ denotes the aggregated feature-distillation term that sums the two AT-based alignments between the teacher and the Simulated-Student at different depths, as described above.  
This configuration allows the teacher to achieve consistent, domain-agnostic representations and multi-scale structural alignment across leads.

\noindent\textbf{Stage 2.}
After the teacher is pretrained, the single-lead student is optimized under the combined supervision of ground-truth labels and the frozen teacher. 
Across the original input and its $K_\mathrm{aug}$ GANs-augmented variants, the student loss is expressed as:
\begin{equation}
\label{eq:student-total-loss-final}
\begin{aligned}
\mathcal{L}_{\text{Student}}
&=
\underbrace{\mathcal{L}_{\text{cls}}}_{\text{diagnostic classification}}
+ \underbrace{0.05\,\mathcal{L}_{\text{Dom}}^{S}}_{\text{domain-adversarial branch}} \\
&\quad
+ \underbrace{0.3\,\widetilde{\mathcal{L}}_{\text{tsld}}^{T\rightarrow S}}_{\text{label distillation}}
+ \underbrace{0.3\,\tilde{\mathcal{L}}_{\text{tsfd}}^{T\rightarrow S}}_{\text{feature distillation}}.
\end{aligned}
\end{equation}

The diagnostic classification loss $\mathcal{L}_{\text{cls}}$ is likewise weighted by 1.0, as accurate supervised learning of rhythm categories remains the primary optimization target guiding the entire distillation process.  
The GRL weight ($0.05$) mirrors that of the teacher stage to maintain consistent domain-invariance pressure.  
Both distillation weights ($0.3$ each) ensure balanced transfer of probabilistic calibration and feature structure from the teacher while preventing overfitting.  
Because the student lacks the hierarchical multi-lead structure of the teacher, a single feature-distillation layer suffices, and multi-view averaging through GANs augmentations broadens its representation space for robust generalization.

The hierarchical weighting allows the teacher to become \emph{self-regularized, domain-agnostic, and multi-scale consistent}, while the student inherits these calibrated representations to achieve robust single-lead AF detection and cross-domain adaptability.

\begin{figure*}[!t]
    
\centering
\includegraphics[scale=0.59]{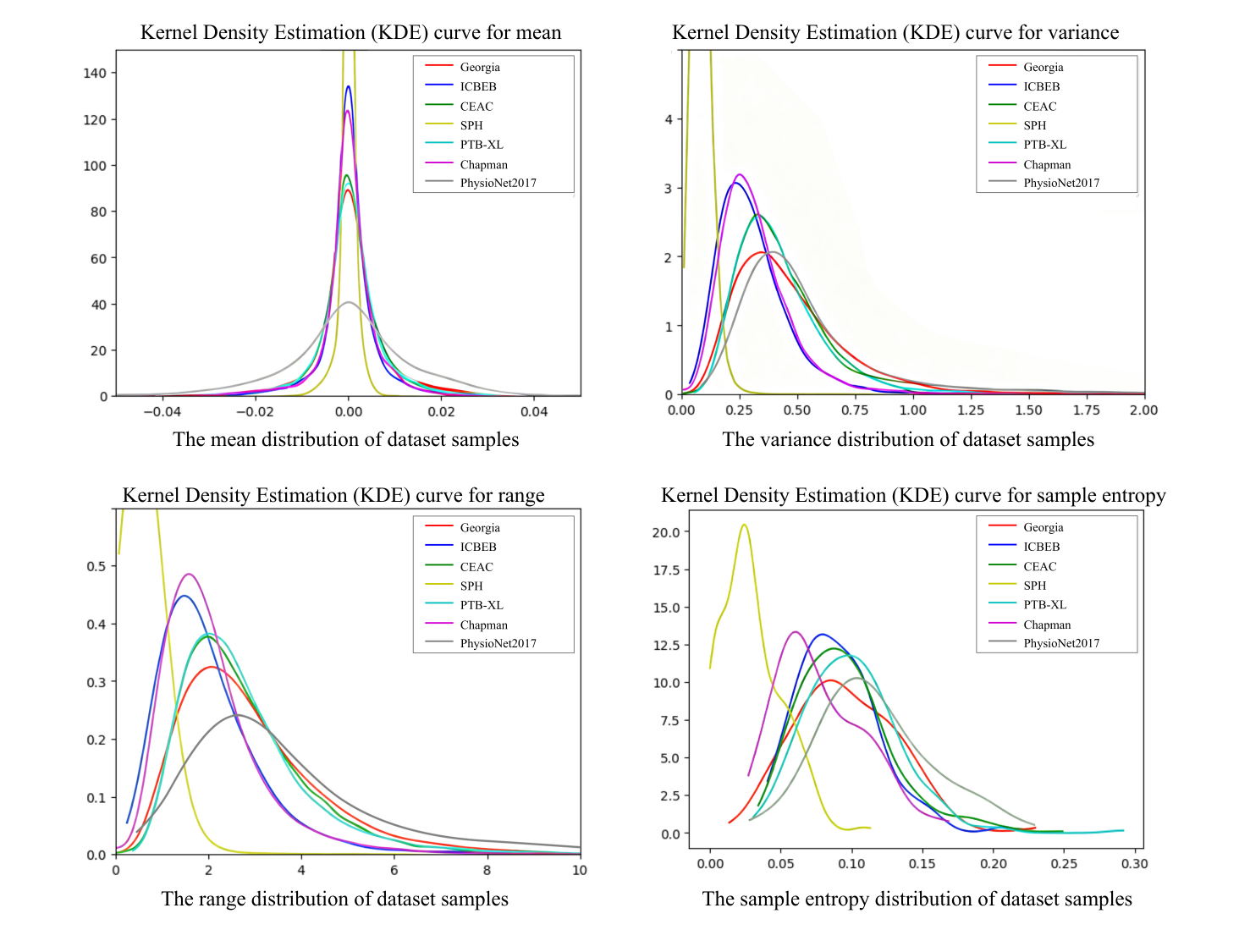}
\caption{KDE-based comparison of sample-level statistics across ECG datasets.
Each curve is one dataset. (a) \textbf{Mean:} Reflects baseline offset and zero-mean consistency.
(b) \textbf{Variance:} Describes amplitude fluctuation; heavier left tails indicate higher energy or variability.
(c) \textbf{Range:} Represents instantaneous dynamic span; larger values indicate more pronounced signal swings.
(d) \textbf{Sample Entropy:} Measures temporal irregularity; right-shifted curves indicate more complex and less periodic rhythms.
}
\label{fig:KDE}
\end{figure*}

\section{Experiment}

\subsection{Dataset}
\noindent\textbf{Ethics and Consent.} This study used only publicly available, de-identified ECG datasets. The original data collection was approved by the corresponding ethics committees/IRBs, and informed consent was obtained or waived by the data providers as described in the original dataset publications. This work involved only secondary analysis and no additional human experiments.

The datasets involved in this study include 6 12-lead open-source ECG datasets (collected by clinical ECG machines): 
PTB-XL\cite{Wagner-2020}, Chapman\cite{Zheng-2020}, Georgia\cite{Alday-2020}, ICBEB\cite{Liu-2018}, CEAC\cite{Ding-2022}, SPH\cite{Zheng-2022}; and 1 single-lead ECG dataset (collected by wearable devices): open-source dataset PhysioNet2017\cite{Clifford2017}.

PTB-XL\cite{Wagner-2020} is a large public 12-lead ECG corpus (greater than 21k 10-s recordings at 100/500 Hz) with rich multi-label clinical annotations (e.g., arrhythmias, MI), widely used for robust ECG classification and anomaly detection. Chapman\cite{Zheng-2020} and Georgia\cite{Alday-2020} are multi-hospital, clinically representative ECG datasets emphasizing arrhythmias, providing single- and multi-lead recordings with diagnostic labels and generally well-standardized durations, thereby supporting large-scale supervised modeling and fair benchmarking. ICBEB\cite{Liu-2018}, CEAC\cite{Ding-2022}, and SPH\cite{Zheng-2022} are diverse, high-quality multi-lead ECG corpora with detailed arrhythmia annotations and mostly fixed-length segments, facilitating consistent preprocessing, robust classification, and anomaly detection across heterogeneous patient populations. PhysioNet2017\cite{Clifford2017} is a public single-lead ECG dataset from multiple clinical sites, released for the 2017 PhysioNet/CinC Challenge, and provides rhythm labels including AF with variable-length recordings, making it a canonical benchmark for arrhythmia detection in wearable-oriented single-lead settings.

\begin{comment}
Self-collected AliveCor (R):To evaluate the generalization performance of the proposed algorithm in real-world applications, we created a self-testing dataset. The data was collected using the same equipment as in the PhysioNet-2017 dataset and includes a total of 19,161 records. All data was clinically collected and annotated by professional online physicians. The dataset comprises 13,762 unannotated records and 5,399 annotated records. In the annotated data, there are 3,548 normal cases, 76 AF cases, 856 other abnormalities, and the remaining records are unidentifiable. The “other abnormalities” category includes over twenty types of arrhythmias, such as tachycardia, bradycardia, ventricular premature beats, and atrial premature beats. Each ECG record contains six types of information: ID, age, gender, height, weight, and diagnostic result.
\end{comment}

The 12-lead ECG datasets all provide standard 12 leads (I, II, III, AVL, AVR, AVF, V1-V6); the single-lead ECG collected by the wearable devices are all I-lead signals. We uniformly process all data into 10s in length. Fig. \ref{fig:KDE} illustrates the inter-dataset differences in baseline, amplitude variability, dynamic range, and temporal complexity using KDE-based statistical feature distributions.

\subsection{Metrics and Experimental Setup}\label{sec:metrics_setup}

\subsubsection{Evaluation Metrics}
We report precision (Pre), recall (Rec), accuracy (Acc), and class-wise F1-scores for the three diagnostic categories:
normal (\(\mathrm{N}\)), atrial fibrillation (\(\mathrm{AF}\)), and other (\(\mathrm{O}\)).
Let \(\mathcal{C}=\{\mathrm{N},\mathrm{AF},\mathrm{O}\}\) denote the set of classes and \(n\) denote the total number of evaluated samples.
For the \(i\)-th sample, \(y_i\in\mathcal{C}\) and \(\hat{y}_i\in\mathcal{C}\) are the ground-truth and predicted labels, respectively.

For a given class \(c\in\mathcal{C}\), we compute \(\mathrm{TP}_c\), \(\mathrm{FP}_c\), \(\mathrm{FN}_c\), and \(\mathrm{TN}_c\) under a one-vs-rest scheme, where
\(\mathrm{TP}_c\) is the number of samples with \(y_i=c\) and \(\hat{y}_i=c\),
\(\mathrm{FP}_c\) counts samples with \(y_i\neq c\) but \(\hat{y}_i=c\),
\(\mathrm{FN}_c\) counts samples with \(y_i=c\) but \(\hat{y}_i\neq c\),
and \(\mathrm{TN}_c\) counts samples with \(y_i\neq c\) and \(\hat{y}_i\neq c\).
Then
\begin{align}
\mathrm{Precision}_c &= \frac{\mathrm{TP}_c}{\mathrm{TP}_c+\mathrm{FP}_c},\\
\mathrm{Recall}_c &= \frac{\mathrm{TP}_c}{\mathrm{TP}_c+\mathrm{FN}_c},\\
\mathrm{F1}_c &= \frac{2\,\mathrm{Precision}_c\,\mathrm{Recall}_c}{\mathrm{Precision}_c+\mathrm{Recall}_c}.
\end{align}

Overall accuracy is computed as the fraction of correctly classified samples:
\begin{equation}
\label{eq:acc}
\mathrm{Acc}
=\frac{1}{n}\sum_{i=1}^{n}\mathbb{I}(\hat{y}_i=y_i)
=\frac{\sum_{c\in\mathcal{C}}\mathrm{TP}_c}{n},
\end{equation}
where \(\mathbb{I}(\cdot)\) is the indicator function that equals \(1\) if the condition holds and \(0\) otherwise.
The equality \(\sum_{c\in\mathcal{C}}\mathrm{TP}_c\) holds because each sample contributes exactly once to the diagonal of the multi-class confusion matrix in the single-label setting. Equivalently, since each sample has exactly one ground-truth label, we have
\(\sum_{c\in\mathcal{C}}(\mathrm{TP}_c+\mathrm{FN}_c)=n\).

The macro-averaged F1, used as our primary summary metric, is the unweighted mean of per-class F1:
\begin{equation}
\label{eq:macro_f1}
\mathrm{F1}_{\mathrm{macro}}=\frac{1}{|\mathcal{C}|}\sum_{c\in\mathcal{C}}\mathrm{F1}_c
=\frac{\mathrm{F1}_{\mathrm{N}}+\mathrm{F1}_{\mathrm{AF}}+\mathrm{F1}_{\mathrm{O}}}{3}.
\end{equation}

\subsubsection{Data Splits and Evaluation Settings}
Unless otherwise stated, training and validation sets are constructed with an 80/20 split (80\% train / 20\% validation per dataset), using subject-level partitioning to prevent patient leakage and stratification to preserve class proportions.

\textbf{1. Real single-lead (Lead~I) test.}
We train and validate on 12-lead ECGs pooled from six open-source datasets with 80/20 splits. Testing is performed on Lead~I recordings from PhysioNet2017 dataset.

\textbf{2. Simulated single-lead (Lead~I) test.}
We perform a leave-one-dataset-out student-transfer protocol on the six 12-lead datasets: in each fold, five datasets are used for Stage-2 student training/validation (again, an 80/20 split per dataset), and Lead~I from the held-out sixth dataset is used exclusively for testing.

\textbf{3. Subject-wise robustness analysis.}
For the real single-lead test, we additionally quantify inter-subject stability by resampling subjects with replacement and recomputing each metric over the selected subjects. This subject-level bootstrap preserves within-subject correlation and estimates 95\% confidence intervals for the final predictions.

\subsubsection{Training Details}

\begin{comment}

For the GAN, 12-lead network (teacher), and student models, we use a batch size of 128.
The learning rate follows a warmup–cosine decay schedule with a maximum learning rate
$\eta_{\max}=10^{-3}$ and a minimum learning rate $\eta_{\min}=10^{-5}$.

The schedule first linearly increases from $0$ to $\eta_{\max}$ during the warmup phase
($t<t_{\mathrm{warm}}$), and then decays to $\eta_{\min}$ following a cosine curve until
the final iteration $T$:

\begin{equation}
\eta(t)=
\begin{array}{@{}l@{}}
t<t_{\mathrm{warm}}\\[6pt]
t\ge t_{\mathrm{warm}}
\end{array}
\left\{
\begin{array}{l}
\eta_{\max}\dfrac{t}{t_{\mathrm{warm}}}\\[7pt]
\begin{aligned}[t]
\eta_{\min} + &\dfrac{1}{2}(\eta_{\max}-\eta_{\min})\\[-2pt]
&\Bigl[\,1 + \cos\!\Bigl(\pi\dfrac{t-t_{\mathrm{warm}}}{T-t_{\mathrm{warm}}}\Bigr)\Bigr]
\end{aligned}
\end{array}
\right.
\label{eq:cosine_schedule}
\end{equation}

The GAN is trained for $600{,}000$ iterations. The teacher model is selected as the
12-lead network checkpoint with the best validation performance, and the student model
used for reporting is the checkpoint with the best validation performance under the same
training setting.
\end{comment}
\noindent\textbf{GANs-based ECG Augmentation.} 
A UNIT-style 1D GANs is trained on six 12-lead ECG datasets using a single RTX~3090 GPU for the real single-lead experiment.
We adopt LSGAN loss, Adam optimization (learning rate $1\times10^{-4}$, betas $(0.5,0.999)$), 
batch size of 1, and a reconstruction-weight factor of 10. 
The trained model produces three style-perturbed single-lead views per mini-batch for Stage~2.

\noindent\textbf{Teacher, Student, and Fine-Tuning.}
All subsequent training is conducted on a single RTX~3090 GPU.

\noindent\textbf{Stage~1.} A 12-lead network is trained on six clinical datasets using batch size~128, 
Adam with a learning-rate schedule from $1\times10^{-3}$ to $1\times10^{-5}$ 
(2 warm-up epochs, 20 total epochs). 
The loss combines classification, domain-adversarial (weight 0.05), 
and self-distillation components. 
The best checkpoint is chosen by the sum of validation F1 across all source domains.

\noindent\textbf{(Stage~2).} A single-lead model (raw + forward-difference channels) is trained on the source datasets specified by each evaluation setting 
with batch size~128 and Adam using the same learning-rate schedule (30 epochs). 
When augmentation is enabled, each mini-batch includes one real view and three GANs-augmented views; for leave-one-dataset-out student evaluation, we set $K_\mathrm{aug}=0$ and use only original views. 
The loss includes classification, domain-adversarial (0.05), 
logit-distillation (0.3), and feature-distillation (0.3) using a queue of size~5.

\noindent\textbf{Fine-Tuning.}
For limited-label scenarios, the Stage~2 model is fine-tuned for 10 epochs 
using only cross-entropy loss, maintaining the same optimizer and schedule, 
with labeling fractions $p \in \{0.03, 0.05, 0.1, 0.15\}$.

\noindent\textbf{Reporting.} Unless noted, all metrics are reported on the test sets defined above; class-wise results (N/AF/O) and $\mathrm{F1}_{\mathrm{macro}}$ (Eq.~\ref{eq:macro_f1}) are provided to reflect performance under class imbalance. For subject-wise robustness, confidence intervals are computed over subjects rather than individual ECG segments.

\noindent\textbf{Label Harmonization (N/AF/O).} The seven ECG datasets provide heterogeneous rhythm/diagnostic annotations (either as rhythm strings or diagnosis codes).
To enable a consistent three-class evaluation, we harmonize all labels into
normal (\(\mathrm{N}\)), atrial fibrillation (\(\mathrm{AF}\)), and other (\(\mathrm{O}\)) using a unified rule set shared across datasets.
Specifically, we assign a record to \(\mathrm{AF}\) if its annotation contains atrial fibrillation or atrial flutter (e.g., AF/AFIB/AFL), since they represent the target atrial tachyarrhythmia screening category in our setting.
We assign a record to \(\mathrm{N}\) only when it is annotated as normal sinus rhythm (e.g., NSR/normal) \emph{without} any additional abnormal rhythm labels.
All remaining rhythms/diagnoses (including non-AF arrhythmias and any non-normal findings) are mapped to \(\mathrm{O}\).

\noindent\textbf{Ambiguous/Overlapping Labels.} For multi-labeled records, we apply a deterministic priority to avoid ambiguity:
(i) if \(\mathrm{AF}\) (or \(\mathrm{AFL}\)) is present, the record is mapped to \(\mathrm{AF}\) regardless of co-occurring labels;
(ii) otherwise, if any non-normal label co-occurs with a normal label, the record is mapped to \(\mathrm{O}\) (i.e., not purely normal).
Records with missing, undefined, or clearly conflicting annotations are excluded from evaluation to ensure label integrity.
The final mapping list was checked for clinical plausibility to ensure that the three-class grouping is medically coherent.

\subsection{Data from Real and Simulation}

\noindent\textbf{Real wearable single-lead evaluation.}
We tested our method on the Lead~I dataset PhysioNet2017 collected by wearable devices and compared it with similar methods. Based on the results in Table~\ref{tab:real_single_lead_physionet2017}, X-LeadGen achieved state-of-the-art (SOTA) performance on the real single-lead test set for AF classification, as measured by $\mathrm{F1}_{\mathrm{macro}}$, $\mathrm{F1}_{\mathrm{AF}}$, $\mathrm{F1}_{\mathrm{N}}$, and accuracy. The gain is most important for $\mathrm{F1}_{\mathrm{AF}}$, because AF screening is the central clinical target of this work rather than broad recognition of heterogeneous non-AF arrhythmias. Although $\mathrm{F1}_{\mathrm{O}}$ is slightly lower than that of SJTU-ECGNet, X-LeadGen provides the best AF-specific and macro-level performance with a moderate parameter budget, supporting its practical use in resource-constrained wearable monitoring.

\begin{table}[htbp]
  \centering
  \caption{Experimental results of real single lead (Lead I) data on different domain generalization ECG methods (Test Set of PhysioNet2017). Macro-F1 follows Eq.~\ref{eq:macro_f1}.}
  \resizebox{0.99\columnwidth}{!}{
    \begin{tabular}{cccccccc}
    \toprule
    \multicolumn{1}{c}{Method} & \multicolumn{1}{c}{$\mathrm{F1}_{\mathrm{macro}}$} & \multicolumn{1}{c}{$\mathrm{F1}_{\mathrm{N}}$} & \multicolumn{1}{c}{$\mathrm{F1}_{\mathrm{AF}}$} & \multicolumn{1}{c}{$\mathrm{F1}_{\mathrm{O}}$} & \multicolumn{1}{c}{$\mathrm{Acc}$} & \multicolumn{1}{c}
    {$\mathrm{Params (M)}$} \\
    \midrule
    
    \multicolumn{1}{c}{ResNet-18 DG\cite{Ballas2022} }      & 0.405 & 0.390  & 0.371  & 0.453  & 0.417  &11.7  \\

    \multicolumn{1}{c}{SE-ResNet AFL\cite{Shang2021}}     & 0.392 & 0.512 & 0.194 & 0.471 & 0.474   &4.0  \\

    \multicolumn{1}{c}{Multi-source ADG\cite{Hasani2020}}      & 0.463 & 0.465 & 0.454 & 0.470  & 0.467  & \underline{0.7}  \\
          
    \multicolumn{1}{c}{GRL\cite{Ganin}}      & 0.601 & 0.566 & 0.738  & 0.499   & 0.554  & \textbf{0.1}  \\
          
    \multicolumn{1}{c}{MBDG\cite{Robey}}      & 0.601 & 0.562 & 0.729  & \underline{0.511} & 0.556  &25.6  \\
          
    \multicolumn{1}{c}{CAT-Net\cite{Islam2024}}      & 0.604 & 0.632 & 0.710  & 0.470 & 0.566  &1.2  \\

    \multicolumn{1}{c}{ST-MEM\cite{Na2024}}      & 0.621 & 0.636 & \underline{0.748} & 0.479 & 0.578  &85.9  \\

    \multicolumn{1}{c}{SJTU-ECGNet\cite{Jin2024}}      & \underline{0.629} & \underline{0.642} & 0.733  & \textbf{0.512} & 0.596  & 1.5 \\

    \multicolumn{1}{c}{BMIRC\cite{Wei2025BMIRC}}      & 0.590 & 0.541 & 0.727  & 0.502 &  \underline{0.600}  &30.1  \\

    \multicolumn{1}{c}{X-LeadGen}      & \textbf{0.646} & \textbf{0.648} & \textbf{0.782} & 0.509 & \textbf{0.607}   &11.5  \\
          
    \bottomrule
    \end{tabular}%
     }
  \label{tab:real_single_lead_physionet2017}%
  
  \vspace{2mm}
  \textbf{Bold} indicates the best performance for each metric, while \underline{underline} denotes the second-best result.

\end{table}%

To directly evaluate across-subject behavior, Table~\ref{tab:subject_wise_robustness} reports a subject-wise bootstrap analysis on the same real wearable test set. Because subjects, rather than individual 10-s ECG segments, are used as the resampling unit, this analysis tests whether the observed performance remains stable under inter-subject variability. X-LeadGen retained the highest macro-F1, AF F1, AF recall, and accuracy under this subject-level evaluation, indicating that the gains in Table~\ref{tab:real_single_lead_physionet2017} were not dominated by a small subset of recordings.

\begin{table*}[t]
  \centering
  \caption{Subject-wise robustness on the real single-lead wearable test set. Values are point estimates with 95\% confidence intervals computed by resampling subjects rather than ECG segments. Macro-F1 follows Eq.~\ref{eq:macro_f1}.}
    \begin{tabular*}{\textwidth}{@{\extracolsep{\fill}}lcccc@{}}
    \toprule
    Method & $\mathrm{F1}_{\mathrm{macro}}$ (95\% CI) & $\mathrm{F1}_{\mathrm{AF}}$ (95\% CI) & AF Recall (95\% CI) & Acc (95\% CI) \\
    \midrule
    SJTU-ECGNet\cite{Jin2024} & \underline{0.629 (0.617--0.641)} & \underline{0.733 (0.711--0.754)} & \underline{0.758 (0.733--0.782)} & 0.596 (0.584--0.608) \\
    BMIRC\cite{Wei2025BMIRC} & 0.590 (0.579--0.601) & 0.727 (0.706--0.748) & 0.744 (0.720--0.767) & \underline{0.600 (0.588--0.612)} \\
    X-LeadGen & \textbf{0.646 (0.635--0.657)} & \textbf{0.782 (0.764--0.799)} & \textbf{0.819 (0.795--0.841)} & \textbf{0.607 (0.595--0.619)} \\
    \bottomrule
    \end{tabular*}
  \label{tab:subject_wise_robustness}
\end{table*}

\noindent\textbf{Simulated single-lead cross-domain evaluation.}
We also conducted a leave-one-dataset-out student-transfer test on six 12-lead open-source datasets. GAN augmentation is disabled in this evaluation ($K_\mathrm{aug}=0$); the augmented setting is used only in the separate wearable single-lead experiment, where no leave-one-dataset-out target holdout is involved. In each fold, one dataset is selected as the test domain, Lead~I is used as the simulated single-lead input, and the remaining five datasets serve as Stage-2 source domains with a five-class domain classifier. Based on Table~\ref{tab:simulated_single_lead_lodo}, X-LeadGen achieved SOTA or near-SOTA AF classification across all held-out domains and obtained the highest average scores over the six target datasets. The improvements are therefore not confined to a single source-target pair; instead, they persist across datasets with different acquisition protocols and patient populations, supporting the robustness of the learned cross-lead and domain-invariant representation.

\noindent\begin{minipage}{\columnwidth}
  \centering
  \refstepcounter{table}
  \label{tab:simulated_single_lead_lodo}
  \noindent\parbox{0.98\columnwidth}{\textbf{Table~\thetable:} Simulated single-lead (Lead I) data experimental results in 6 datasets.}
  \vspace{1mm}
  \resizebox{0.99\columnwidth}{!}{
    \begin{tabular}{ccccccc}
    \toprule
    \multicolumn{1}{c}{Test Set} & Method & \multicolumn{1}{c}{$\mathrm{F1}_{\mathrm{macro}}$} & \multicolumn{1}{c}{$\mathrm{F1}_{\mathrm{N}}$} & \multicolumn{1}{c}{$\mathrm{F1}_{\mathrm{AF}}$} & \multicolumn{1}{c}{$\mathrm{F1}_{\mathrm{O}}$} & \multicolumn{1}{c}{$\mathrm{Acc}$} \\
    \midrule
    \multicolumn{1}{c}{\multirow{4}[2]{*}{Georgia}} & ST-MEM\cite{Na2024} & 0.691 & 0.530 & 0.710 & 0.833 & 0.736 \\
          & SJTU-ECGNet\cite{Jin2024} & 0.694 & 0.537  & 0.703 & \textbf{0.842} & \underline{0.767} \\
          & BMIRC\cite{Wei2025BMIRC} & \textbf{0.708} & \textbf{0.556} & \underline{0.729} & \underline{0.839} & \textbf{0.769} \\
          & X-LeadGen  & \underline{0.706} & \underline{0.549} & \textbf{0.731} & 0.837 & 0.766 \\
    \midrule
    \multicolumn{1}{c}{\multirow{4}[2]{*}{ICBEB}} & ST-MEM\cite{Na2024} & \underline{0.748} & \underline{0.548} & 0.867  & \textbf{0.830} & \textbf{0.782} \\
          & SJTU-ECGNet\cite{Jin2024} & 0.747  & \underline{0.548}  & 0.865 & \underline{0.829} & 0.780 \\
          & BMIRC\cite{Wei2025BMIRC} & 0.744 & 0.546 & \underline{0.869} & 0.817  & 0.770 \\
          & X-LeadGen  & \textbf{0.755} & \textbf{0.565} & \textbf{0.873} & 0.827 & \underline{0.781} \\
    \midrule
    \multicolumn{1}{c}{\multirow{4}[2]{*}{CEAC}} & ST-MEM\cite{Na2024} & 0.803 & 0.673 & 0.935 & 0.801 & 0.771 \\
          & SJTU-ECGNet\cite{Jin2024} & 0.793 & 0.661 & 0.931 & 0.787  & 0.757 \\
          & BMIRC\cite{Wei2025BMIRC} & \underline{0.831} & \underline{0.716}  & \textbf{0.950} & \underline{0.828} & \underline{0.801} \\
          & X-LeadGen  & \textbf{0.843} & \textbf{0.734} & \underline{0.943}  & \textbf{0.853} & \textbf{0.824} \\
    \midrule
    \multicolumn{1}{c}{\multirow{4}[2]{*}{SPH}} & ST-MEM\cite{Na2024} & 0.751 & 0.657  & 0.881 & 0.715 & 0.685 \\
          & SJTU-ECGNet\cite{Jin2024} & 0.708 & 0.595 & 0.849 & 0.679 & 0.649 \\
          & BMIRC\cite{Wei2025BMIRC} & \textbf{0.808} & \textbf{0.766} & \textbf{0.924} & \textbf{0.733} & \textbf{0.755} \\
          & X-LeadGen  & \underline{0.785}  & \underline{0.726}  & \underline{0.909} & \underline{0.720}  & \underline{0.728} \\
    \midrule
    \multicolumn{1}{c}{\multirow{4}[2]{*}{PTB-XL}} & ST-MEM\cite{Na2024} & 0.576 & 0.540 & 0.858  & \underline{0.329} & 0.479 \\
          & SJTU-ECGNet\cite{Jin2024} & 0.572 & 0.542 & 0.855 & 0.320 & 0.479 \\
          & BMIRC\cite{Wei2025BMIRC} & \underline{0.586} & \underline{0.557}  & \underline{0.874}  & 0.327  & \underline{0.494} \\
          & X-LeadGen  & \textbf{0.618} & \textbf{0.605} & \textbf{0.899} & \textbf{0.349} & \textbf{0.535} \\
    \midrule
    \multicolumn{1}{c}{\multirow{4}[2]{*}{Chapman}} & ST-MEM\cite{Na2024} & 0.853 & 0.768  & \underline{0.882}  & 0.909 & 0.882 \\
          & SJTU-ECGNet\cite{Jin2024} & 0.846 & 0.752 & 0.878  & 0.907  & 0.878 \\
          & BMIRC\cite{Wei2025BMIRC} & \underline{0.872} & \textbf{0.829} & 0.868 & \underline{0.919}  & \underline{0.895} \\
          & X-LeadGen  & \textbf{0.884} & \underline{0.813} & \textbf{0.911} & \textbf{0.928} & \textbf{0.906} \\
    \midrule
    \multicolumn{1}{c}{\multirow{4}[2]{*}{Avg}} & ST-MEM\cite{Na2024} & 0.737 & 0.619 & 0.856 & 0.736 & 0.723 \\
          & SJTU-ECGNet\cite{Jin2024} & 0.727 & 0.606 & 0.847 & 0.727  & 0.718 \\
          & BMIRC\cite{Wei2025BMIRC} & \underline{0.758}  & \underline{0.662} & \underline{0.869} & \underline{0.744} & \underline{0.747} \\
          & X-LeadGen  & \textbf{0.765} & \textbf{0.665} & \textbf{0.878} & \textbf{0.752} & \textbf{0.757} \\
    \bottomrule
    \end{tabular}
    }

  \vspace{2mm}
  \textbf{Bold} indicates the best performance for each metric, while \underline{underline} denotes the second-best result.
\end{minipage}\par

Table~\ref{tab:cross_domain_stability} further summarizes fold-level stability across the six leave-one-dataset-out tests. The mean and standard deviation quantify average performance and variability across target domains, while the top-2 count tests whether a method remains competitive across datasets rather than succeeding on only one favorable domain. X-LeadGen obtained the highest mean $\mathrm{F1}_{\mathrm{macro}}$, $\mathrm{F1}_{\mathrm{AF}}$, and accuracy, and ranked first or second by $\mathrm{F1}_{\mathrm{macro}}$ in all six folds. This compact analysis shows that the improvement is distributed across held-out domains rather than driven by a single dataset.

\noindent\begin{minipage}{\columnwidth}
  \centering
  \refstepcounter{table}
  \label{tab:cross_domain_stability}
  \noindent\textbf{Table~\thetable:} Cross-domain stability across six leave-one-dataset-out folds.
  Top-2 folds denotes the number of target domains in which each method ranks first or second by $\mathrm{F1}_{\mathrm{macro}}$.
  \par\vspace{1mm}
  \resizebox{0.99\columnwidth}{!}{
    \begin{tabular}{ccccc}
    \toprule
    Method & $\mathrm{F1}_{\mathrm{macro}}$ mean$\pm$SD & $\mathrm{F1}_{\mathrm{AF}}$ mean$\pm$SD & Acc mean$\pm$SD & Top-2 folds \\
    \midrule
    ST-MEM\cite{Na2024} & 0.737$\pm$0.096 & 0.856$\pm$0.076 & 0.723$\pm$0.136 & 1/6 \\
    SJTU-ECGNet\cite{Jin2024} & 0.727$\pm$0.094 & 0.847$\pm$0.076 & 0.718$\pm$0.138 & 0/6 \\
    BMIRC\cite{Wei2025BMIRC} & \underline{0.758$\pm$0.103} & \underline{0.869$\pm$0.076} & \underline{0.747$\pm$0.134} & \underline{5/6} \\
    X-LeadGen & \textbf{0.765$\pm$0.096} & \textbf{0.878$\pm$0.075} & \textbf{0.757$\pm$0.124} & \textbf{6/6} \\
    \bottomrule
    \end{tabular}
  }
\end{minipage}\par

\begin{figure*}[!t]
  \centering
  \includegraphics[width=0.9\textwidth]{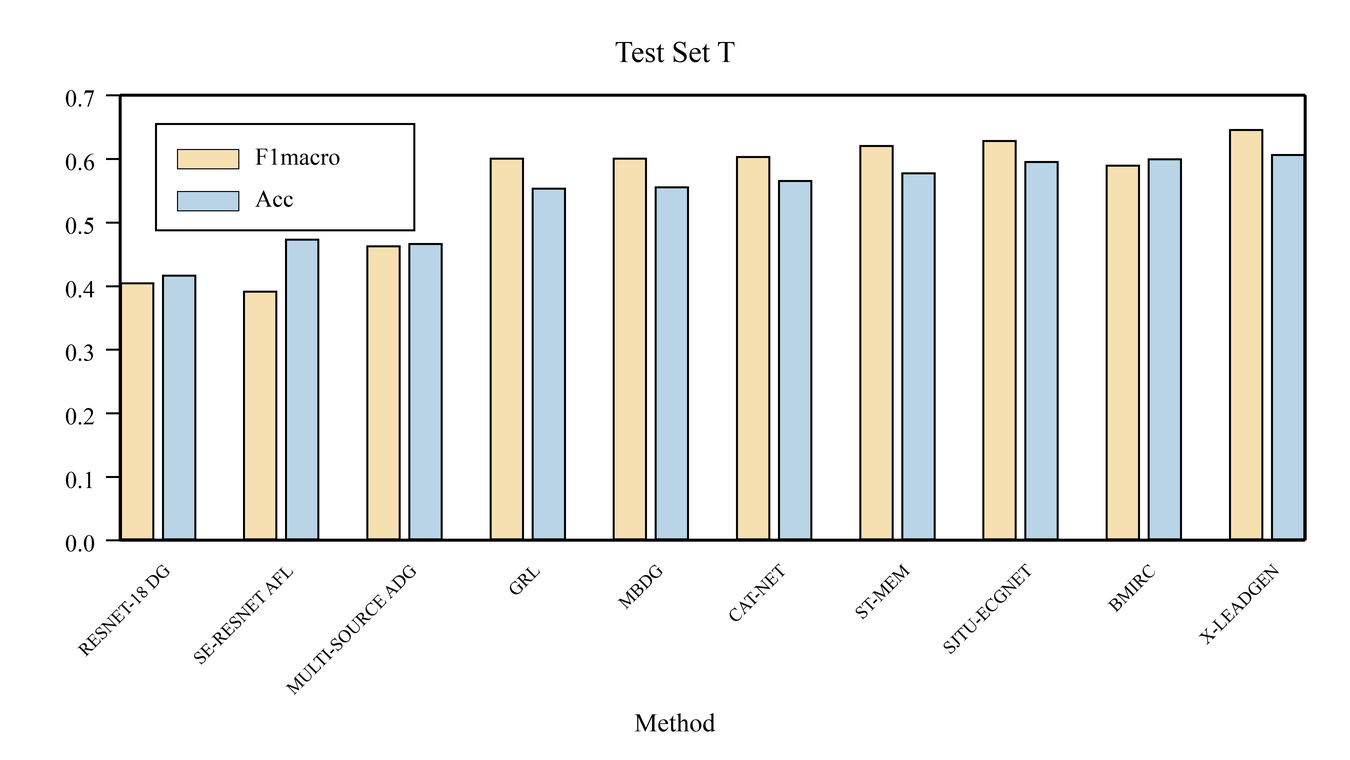}
  \caption{Comparison of $\mathrm{F1}_{\mathrm{macro}}$ and Accuracy of different domain generalization methods on PhysioNet2017 dataset.}
  \label{fig:real_single_lead_bars}
\end{figure*}

\subsection{Ablation Study}

\noindent\textbf{Experimental Setup.}  
Ablation experiments were conducted on the real single-lead dataset (Test Set~PhysioNet2017) using six module combinations (Table~\ref{tab:ablation_modules}).  
Four components were examined: (1) \textit{Student Domain Adversarial (GRL)}, (2) \textit{Teacher–Student Distillation}, (3) \textit{Teacher Self-Distillation}, and (4) \textit{Data Augmentation} (GANs synthesis and forward-difference).  
The models were: \textbf{M1:} Student Baseline; \textbf{M2:} Domain Adversarial only; \textbf{M3:} +teacher–student distillation; \textbf{M4:} +teacher self-distillation; \textbf{M5:} +data augmentation (exclude distillation); and \textbf{M6:} full model.

\begin{table}[htbp]
  \centering
  \tabcolsep 0.1cm
  \caption{Module configurations for the ablation study.}
  \label{tab:ablation_modules}
    \begin{tabular}{p{13em}ccccccc}
    \toprule
    Module & \multicolumn{1}{c}{1} & \multicolumn{1}{c}{2} & \multicolumn{1}{c}{3} & \multicolumn{1}{c}{4} & \multicolumn{1}{c}{5} & \multicolumn{1}{c}{6} \\ 
    \midrule
    Student Domain Adversarial & ×     & $\surd$    & $\surd$      & $\surd$     & $\surd$     & $\surd$ \\
    Teacher-Student Distillation & ×     & ×     & $\surd$        & $\surd$     & ×     & $\surd$ \\
    Teacher Self-Distillation & ×     & ×     & ×     & $\surd$     & ×     & $\surd$ \\
 
    Data Augmentation & ×     & ×     & ×     & ×     & $\surd$     & $\surd$ \\
    \bottomrule
    \end{tabular}%

\end{table}%

\begin{figure}[htbp]
\centering

\includegraphics[width=\columnwidth]{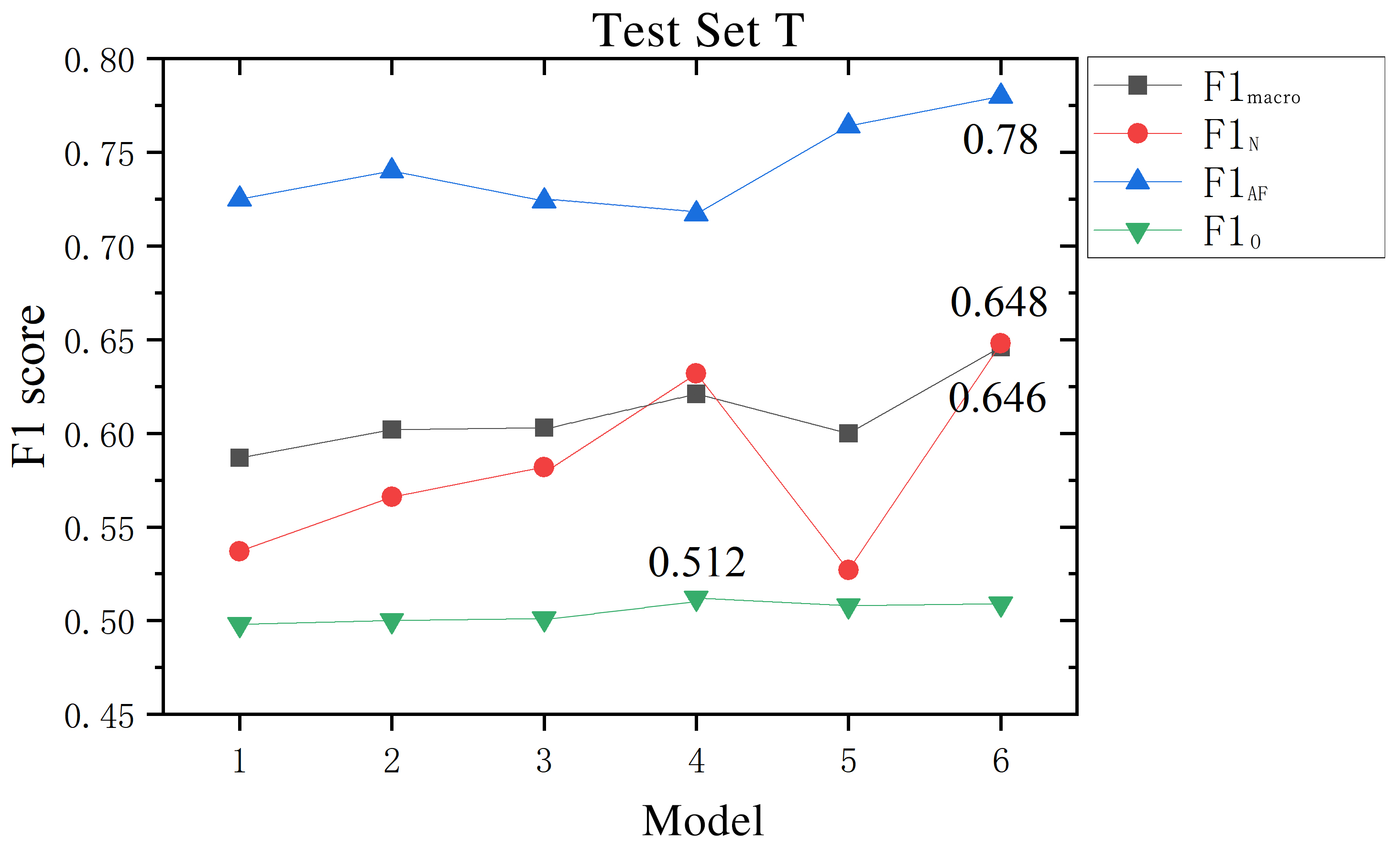}
\caption{Performance comparison in terms of $\mathrm{F1}$ scores for various module configurations on the Test Set~T.} 
\label{fig:ablation_curve} 
\end{figure}

\noindent\textbf{Results and Trend.}  
As shown in \textbf{Fig. \ref{fig:ablation_curve}}, performance steadily improves as modules are added.  
\textbf{M2} shows that domain adversarial learning alone significantly enhances cross-domain robustness.  
\textbf{M3--M4} yield further gains, confirming the stabilizing role of teacher-guided distillation.  
\textbf{M5} improves moderately but underperforms the distillation-based setups.  
The strongest overall setting is \textbf{M6}, achieving the best $\mathrm{F1}_{\mathrm{macro}} = 0.646$ ($\mathrm{F1}_{\mathrm{AF}} = 0.782$, $\mathrm{F1}_{\mathrm{N}} = 0.648$, $\mathrm{F1}_{\mathrm{O}} = 0.509$).

\noindent\textbf{Discussion.}  
The results reveal that multi-lead knowledge transfer and domain adversarial learning act synergistically.  
The teacher network provides multi-scale morphological and rhythm features, while the domain adversarial training branch mitigates dataset bias.  
Data augmentation adds signal diversity and complements the other modules, but augmentation alone does not replace teacher-guided representation transfer.  
This pattern explains the robustness observed in the real wearable and leave-one-dataset-out evaluations: X-LeadGen jointly reduces dataset-specific bias, preserves AF-related rhythm evidence, and exposes the student to realistic single-lead variability.

\subsection{Small-Sample Fine-Tuning}

To further explore the effectiveness of fine-tuning our framework under limited supervision, we conducted small-sample fine-tuning experiments. The model was initially pre-trained on six large open-source datasets containing 12-lead ECG signals, followed by fine-tuning on a real-world wearable dataset PhysioNet2017 using varying proportions of labeled data (\(3\%\), \(5\%\), \(10\%\), and \(15\%\)), as summarized in Table~\ref{tab:small_sample_finetune}. 

\begin{table}[htbp]
  \centering
  \caption{Small-sample fine-tuning on PhysioNet2017: absolute $\mathrm{F1}$ and relative changes (\%) from the non-fine-tuned \textbf{M6} baseline across label ratios (0\%, 3\%, 5\%, 10\%, 15\%). The \textit{None} column reports the baseline score and the average relative change over the four fine-tuning ratios.}

  \resizebox{\columnwidth}{!}{%
    \begin{tabular}{cccccc}
    \toprule
    \multicolumn{1}{c}{Category} & \multicolumn{1}{c}{None} & 3\%   & 5\%   & 10\%  & 15\% \\
    \midrule
    \multirow{2}[2]{*}{$\mathrm{F1}_{\mathrm{macro}}$} & 0.646 & 0.701 & 0.733 & 0.758 & 0.757 \\
          & +14.13\%(AVG) & +8.51\% & +13.47\% & \textbf{+17.34\%} & +17.18\% \\
    \midrule
    \multirow{2}[2]{*}{$\mathrm{F1}_{\mathrm{N}}$} & 0.648 & 0.788 & 0.841 & 0.872 & 0.871 \\
          & +30.09\%(AVG) & +21.60\% & +29.78\% & \textbf{+34.57\%} & +34.41\% \\
    \midrule
    \multirow{2}[2]{*}{$\mathrm{F1}_{\mathrm{AF}}$} & 0.782  & 0.729 & 0.735 & 0.781 & 0.771 \\
          & -3.58\%(AVG) & -6.78\% & -6.01\% & \textbf{-0.13\%} & -1.41\% \\
    \midrule
    \multirow{2}[2]{*}{$\mathrm{F1}_{\mathrm{O}}$} & 0.509 & 0.585 & 0.621 & 0.620  & 0.632 \\
          & +20.73\%(AVG) & +14.93\% & +22.00\% & +21.81\% & \textbf{+24.17\%} \\
    \bottomrule
    \end{tabular}%
  }
  \label{tab:small_sample_finetune}%
\end{table}

In Table~\ref{tab:small_sample_finetune}, the \textit{AVG} percentages in the \textit{None} column denote, for each F1 metric, the mean relative change (increase or decrease) achieved by fine-tuning with \(3\%\), \(5\%\), \(10\%\), and \(15\%\) labeled data, compared with the non-fine-tuned baseline. The small-sample fine-tuning experiments show that $\mathrm{F1}_{\mathrm{macro}}$, $\mathrm{F1}_{\mathrm{N}}$, and $\mathrm{F1}_{\mathrm{O}}$ improve even with limited labeled data, with gains that gradually saturate as supervision increases. This indicates that multi-lead pretraining and domain-invariant representations already furnish a strong initialization, and further labels mainly yield incremental refinements. A notable exception is \(\mathrm{F1}_{\mathrm{AF}}\), which remains close to the baseline at \(10\%\) but decreases at the other label ratios, suggesting that AF-specific features are largely saturated by pretraining and can be degraded by small, noisy target sets. By contrast, \(\mathrm{F1}_{\mathrm{N}}\) and \(\mathrm{F1}_{\mathrm{O}}\) increase steadily with more labels, benefiting from target-domain calibration. Overall, these results underscore a key strength of our framework: robust AF detection is already achievable without fine-tuning, making the pre-trained model particularly well-suited to wearable scenarios with scarce labeled data.

\section{Limitation}
Despite the promising results, several limitations remain. First, although we evaluated the framework on six heterogeneous 12-lead datasets and one wearable single-lead dataset, these sources still underrepresent the diversity of devices, noise conditions, patient populations, and care settings encountered in large-scale deployment; broader validation should therefore include multi-device and longitudinal wearable cohorts. Second, our knowledge transfer pipeline assumes access to multi-lead data during pretraining. When only single-lead signals are available, future work could explore self-supervised or contrastive pretraining and structure-aware constraints inspired by recent neural signal studies on functional decoupling, adaptive cognitive graphs, cognitive-state dynamics, and hyperbolic prototype learning~\cite{Li2026AdaptiveDecoupling,Gao2025AdaptiveCognitiveGraphs,Li2023BCDRN,Gao2026HyperbolicPrototype}. Third, X-LeadGen models ECG alone; multimodal physiological extensions that combine cardiac dynamics with neural or attention-related signals may further improve robustness under affective, cognitive-load, and free-living conditions~\cite{Gao2026BrainHeart,Gao2025DynamicAuditory,Ye2023CGRUAttention}. These directions should be evaluated with subject-independent protocols and carefully controlled computational budgets before clinical deployment.

\section{Conclusion}

In this work, we presented \textbf{X-LeadGen}, a two-stage, multi-lead–guided and domain-invariant framework for AF detection from single-lead wearable ECG. By combining \textit{multi-scale distillation}, \textit{domain-adversarial alignment}, and \textit{physiology-aware data augmentation}, X-LeadGen transfers both morphological and rhythm information from 12-lead clinical ECGs to single-lead deployments. Experiments on seven real-world ECG datasets show consistent gains over state-of-the-art domain generalization baselines, while subject-wise, leave-one-dataset-out, ablation, and small-sample fine-tuning analyses confirm the robustness, complementarity, and data efficiency of the pre-trained model. More broadly, X-LeadGen provides a scalable recipe for transferring diagnostic capability from clinical-grade multi-lead systems to resource-constrained wearables and for cross-domain knowledge transfer in physiological signal analysis toward lightweight, reliable, and generalizable AI-based cardiac monitoring.

\begin{comment}
    
\appendices

Appendixes, if needed, appear before the acknowledgment.

\section*{Acknowledgment}

The preferred spelling of the word ``acknowledgment'' in American English is 
without an ``e'' after the ``g.'' Use the singular heading even if you have 
many acknowledgments. Avoid expressions such as ``One of us (S.B.A.) would 
like to thank $\ldots$ .'' Instead, write ``F. A. Author thanks $\ldots$ .'' In most 
cases, sponsor and financial support acknowledgments are placed in the 
unnumbered footnote on the first page, not here.

\end{comment}

\clearpage
\setcounter{section}{0}
\setcounter{subsection}{0}
\setcounter{figure}{0}
\setcounter{table}{0}
\setcounter{equation}{0}
\renewcommand{\thesection}{S\arabic{section}}
\renewcommand{\thesubsection}{\thesection.\arabic{subsection}}
\renewcommand{\thefigure}{S\arabic{figure}}
\renewcommand{\thetable}{S\arabic{table}}
\renewcommand{\theequation}{S\arabic{equation}}
\renewcommand{\theHsection}{supp.\arabic{section}}
\renewcommand{\theHfigure}{supp.\arabic{figure}}
\renewcommand{\theHtable}{supp.\arabic{table}}
\renewcommand{\theHequation}{supp.\arabic{equation}}

\section*{Supplementary Materials}

\section{Detailed Architecture of the FibRes Backbone}
\label{sec:supp_fibres_backbone}

The teacher and student networks in X-LeadGen share the same FibRes backbone, while accepting 12-lead and single-lead ECG inputs, respectively. As illustrated in Supplementary Fig.~\ref{fig:supp_fibres_backbone}, the input ECG is first projected by a one-dimensional convolution with a kernel size of 16 and a stride of 1. This stem layer converts the waveform into an initial feature representation that is subsequently processed by 16 stacked Fibrillation Residual (FibRes) Blocks.

Each FibRes Block begins with max pooling, which reduces the temporal resolution before the convolutional transformations. The pooled features are then processed sequentially by batch normalization, a one-dimensional convolution, a second batch-normalization layer, ReLU activation, dropout, and a second one-dimensional convolution. In parallel, the pooled features are propagated through a shortcut connection. When the feature dimensions of the two paths differ, the shortcut is adjusted by channel alignment before the two paths are combined through element-wise residual addition. This organization preserves a direct information path across the block while allowing the main branch to learn transformations from a temporally compressed representation.

Stacking 16 FibRes Blocks progressively converts the input waveform into a compact high-level representation. Max pooling provides hierarchical temporal abstraction, whereas the residual connections retain information from the preceding representation at every stage. Channel alignment makes the residual addition well defined when the feature width changes across successive groups of blocks. The output of the final FibRes Block is passed to the Feature Aggregation module described in the main text, which produces the representation used by the downstream diagnosis and domain-adversarial branches.

\begin{figure}[t]
    \centering
    \includegraphics[width=0.68\columnwidth]{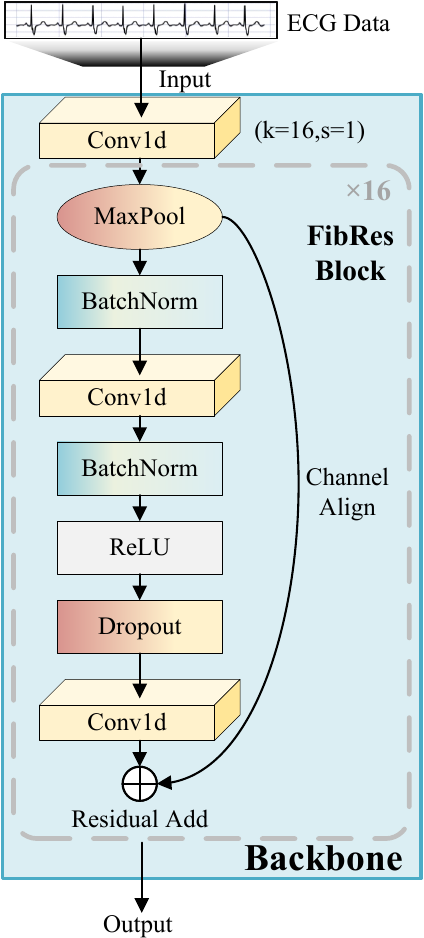}
    \caption{Detailed architecture of the FibRes backbone used in X-LeadGen. The input ECG is first processed by a one-dimensional convolution with a kernel size of 16 and a stride of 1, followed by 16 stacked FibRes Blocks. Each block applies max pooling before the main convolutional transformations. The pooled representation is also propagated through a shortcut path, where channel alignment is applied when required, and is combined with the transformed representation by residual addition.}
    \label{fig:supp_fibres_backbone}
\end{figure}

\section{Additional Limitations}
\label{sec:supp_limitations}

References from image, audio, 3D, and general artificial-intelligence research are used below only as cross-domain methodological precedents, not as evidence that the corresponding techniques have been validated for ECG analysis.

First, the current GAN-based augmentation is designed to broaden the training distribution, but it does not provide explicit control over which waveform characteristics are generated, retained, or suppressed. In visual generative systems, concept removal has been approached through parameter editing, trajectory steering, and rectified-flow interventions, while subsequent work has shown that apparently erased concepts can sometimes be recovered through optimized generation trajectories~\cite{lu2024mace,li2025set,gao2024eraseanything,gao2025revoking}. These studies expose a relevant but untested question for synthetic ECG: whether an augmentation model can inadvertently remove, attenuate, or reintroduce diagnostically important morphology under changes in sampling conditions. X-LeadGen currently neither enforces explicit physiological constraints on generated waveforms nor evaluates targeted suppression and recovery of AF-related patterns. Consequently, augmented samples should be interpreted as a training aid rather than as independently validated physiological records.

Second, the present study does not establish provenance, tamper resistance, or demographic fairness for synthesized ECG segments. Adjacent research has examined watermark survival under generative image editing, post-hoc detection of copy--move manipulation, and key-secured information hiding in 3D representations~\cite{lu2024robust,lu2022copy,ren2025all}; these lines of work address complementary aspects of traceability, forensic detection, and controlled access, but have not yet been adapted to ECG generation. More broadly, analyses of bias in generative AI and integrated AI-governance frameworks emphasize that reliability, fairness, transparency, and accountability must be evaluated jointly rather than inferred from task accuracy alone~\cite{lin2025bias,jiang2026never}. X-LeadGen does not yet embed sample-level provenance, test whether synthetic data alter performance across demographic or acquisition subgroups, or evaluate adversarial manipulation of the augmentation pipeline. These omissions limit claims about trustworthy reuse of the generated data beyond the controlled experiments reported here.

\section{Future Research Directions}
\label{sec:supp_future_work}

A first direction is to develop controllable, physiology-constrained ECG generation and editing. Training-free cross-domain composition, physically plausible image composition with pretrained flow models, and region-supervised drag editing provide distinct examples of how a pretrained generative prior can be steered while preserving surrounding content~\cite{lu2023tf,lu2025does,zhou2025dragflow}. A signal-domain analogue could operate on selected temporal intervals or waveform components while constraining rhythm, morphology, and inter-beat continuity. Such a system would make it possible to vary nuisance factors without corrupting AF evidence and to test explicitly which generated changes improve cross-device generalization. This proposal is methodological: the cited visual techniques do not themselves establish physiological validity.

A second direction is to enrich the representation learned before and during cross-lead distillation. Directly relevant ECG studies suggest combining contrastive and reconstructive self-supervision and extending 12-to-1-lead knowledge transfer to related rhythm disorders~\cite{qin2026coreecg,zhou2026slld}. Complementary work in event-based recognition and general time-series analysis highlights explicit long-range temporal modeling, differencing-based contrastive learning, and joint time--frequency consistency~\cite{yang2025temporal,zhang2025dconad,zhang2025frect}. Cross-domain architectures based on hierarchical frequency fusion and wavelet-domain parameter-efficient adaptation further suggest ways to preserve multiscale structure without fully retraining a large encoder~\cite{lin2025butter,xuan2026wavesp}. For X-LeadGen, these ideas motivate ECG-specific pretraining and adaptive temporal--frequency modules, but their value should be established through ablation studies under subject-independent and leave-one-dataset-out protocols.

A third direction is to make knowledge transfer more selective and computationally efficient. Beyond conventional logit matching, other application domains have reused intermediate attention for relational reasoning and combined synthetic examples with domain-adaptive distillation for compact models~\cite{li2025reusing,wu2026squadsql}. Work on robust human--scene contact perception also illustrates the transfer of complementary pretrained priors together with uncertainty-oriented training under occlusion and noise~\cite{lin2026graphicontact}. These precedents motivate lead-aware teacher selection, uncertainty-weighted feature matching, and reuse of intermediate attention maps in X-LeadGen, particularly when the available leads, labels, or computational budgets vary across devices.

A fourth direction is to move from a single-signal classifier toward multimodal and decentralized learning. Anchor-based multimodal embedding offers one example of synchronizing heterogeneous representations, while recent cardiovascular analyses identify opportunities and unresolved clinical challenges for large ECG and imaging models~\cite{shen2025mess,chen2026cardiovascular}. At the population level, heterophily-aware message passing and federated graph-level clustering provide methodological starting points for modeling relationships among patients, devices, and institutions without assuming that connected domains are statistically similar~\cite{zhang2025dual,zhang2026dynamic}. Future studies could therefore combine ECG with imaging, photoplethysmography, or clinical variables and evaluate privacy-preserving collaboration across centers; these extensions require prospective validation and cannot be inferred from the present single-modality experiments.

Finally, future work should improve optimization diagnostics and reproducibility while retaining human control over clinical model selection. Execution-grounded machine-learning engineering agents show how iterative debugging and search can be evaluated in verifiable environments~\cite{yang2026frontisma1}, whereas parametric anchoring exemplifies the stabilization of difficult numerical iterations~\cite{lin2024parametric}. These ideas motivate constraint-aware search and convergence diagnostics for X-LeadGen's coupled adversarial, generative, and distillation objectives. Automated optimization must remain subordinate to prespecified patient-level validation, physiological plausibility checks, and expert review.

\end{document}